\documentclass{PoS}
\usepackage{graphicx}% Include figure files
\usepackage{enumerate} %%
\usepackage{amsmath,amssymb}

\title{
\begin{picture}(0,0)(0,0)%
\put(355,95){\makebox(0,0)[l]{\textnormal{\normalsize 
YITP-07-75}}}%
\put(355,85){\makebox(0,0)[l]{\textnormal{\normalsize 
KUNS-2106}}}%
\put(355,75){\makebox(0,0)[l]{\textnormal{\normalsize 
KEK-CP-204}}}%
\end{picture}%
High precision study of $B^*B\pi$ coupling \\
in unquenched QCD}

\ShortTitle{High precision study of $B*B\pi$ coupling in unquenched QCD}

\author{\speaker{Hiroshi Ohki} 
\\
       Yukawa Institute for Theoretical Physics, Kyoto
       University, Kyoto 606-8502, Japan,\\
        Department of Physics, Kyoto University, Kyoto 606-8501, Japan \\
        E-mail: \email{ohki@yukawa.kyoto-u.ac.jp}}
\author{Hideo Matsufuru  \\
        High Energy Accelerator Research Organization(KEK), Tsukuba
        305-0801, Japan \\
        E-mail: \email{hideo.matsufuru@kek.jp}}
\author{Tetsuya Onogi  \\
        Yukawa Institute for Theoretical Physics, Kyoto University,
        Kyoto 606-8502, Japan\\
        E-mail: \email{onogi@yukawa.kyoto-u.ac.jp}}

\abstract{
The $B^* B\pi$ coupling is a fundamental parameter of chiral 
effective Lagrangian with heavy-light mesons and can constrain 
the $B\rightarrow \pi l \nu$ form factor in the soft pion limit
which will be useful for precise determination of $|V_{ub}|$. 
We compute the $B^* B \pi $ coupling 
with the static heavy quark and the $O(a)$-improved Wilson light
quark.
Simulations  are carried out with $n_f=2$ unquenched $12^3\times 24$
lattices at $\beta=1.80$ generated by CP-PACS collaboration.
Following the quenched study by Negishi et al.,
we employ the all-to-all propagator with 200 low eigenmodes as well 
as HYP smeared link to improve the statistical accuracy.
}

\FullConference{The XXV International Symposium on 
Lattice Field Theory\\
		 July 30-4 August 2007\\
		 Regensburg, Germany}

\begin{document}

\section{Introduction}

One of the major subjects in particle physics is to determine
the CKM matrix elements in order to test the standard model and find a clue to
 the physics beyond.
Among all the components, $|V_{ub}|$ is very attractive,
since it is determined from the electro-weak (EW) 
tree-level processes and hence gives a good reference point for the
 standard model  prediction. 

Despite its importance, $|V_{ub}|$ is known only with 10\% accuracy 
from inclusive decay and 20-30\% accuracy from exclusive decay.
Although the inclusive determination has been developed for the last
decades, 
it suffers from sizable higher order correction in OPE 
due to the limited kinematic range to avoid the background 
from $B\rightarrow X_c l\nu$. 
In order to reduce the error, significant theoretical
improvement is needed.
On the other hand, the determination from 
the exclusive $B\rightarrow \pi l\nu$ decay is more promising, 
owing to the significant progress in the experiment. 
Recently BABAR collaboration gave a very precise  
determination of the form factor up to $|V_{ub}|$~\cite{0612020}. 
The only problem is the theoretical uncertainty in the form factor.
In view of the present and future experimental situation, 
in order to determine $|V_{ub}|$ 
one only needs to know the  form factor at any single point 
in $q^2$ in principle. 
In practice, what would be the best choice of $q^2$?  
Normally one can choose the $q^2$ for 
the smallest nonzero recoil momenta.  
However, there is an alternative choice of $q^2 \sim m^2_B$, 
where one can use the symmetry relation.
In this choice the $B \to \pi l \nu$ form factor 
$f^+(q^2)$ can be expressed
in terms of the $B^*$ meson decay constant $f_{B^*}$ and 
the $B^* B\pi$ coupling $\hat{g}_b$ as 
\begin{align}
f^+(q^2)
=-\frac{f_{B^*}}{2f_\pi}
\left[
\hat{g}_b
\left(
\frac{m_{B^*}}{v \cdot k-\Delta}-\frac{m_{B^*}}{m_B}
\right)
+\frac{f_B}{f_{B^*}}
\right],
\end{align}
where $v$ is the velocity of the $B$ meson,
$k$ is the pion momentum and  $\Delta=m_{B^*}-m_{B}$.
Therefore one can reduce problems of computing the form factor to
simpler problems of computing the decay constant and effective 
coupling. Since the $B^*$ state is below the $B\pi$ threshold, 
it is not possible to measure the $B^*\rightarrow B\pi$ decay 
experimentally and only lattice QCD can provide precise information
of the $B^* B\pi$ coupling. 

In lattice simulation,
the heavy-light decay constant
has been studied extensively and is expected to be determined 
precisely with nonperturbative accuracy in unquenched QCD near
future. However, there has not been much progress in the 
study of the  $B^* B \pi$ coupling, in particular for unquenched QCD.
Therefore, in this report we present our recent study of high
precision determination of the $B^* B \pi$ coupling in unquenched QCD.

At present, one of the promising approaches to computing the $B^* B\pi$
coupling is to use the nonperturbative 
heavy quark effective theory(HQET) including $1/M$ corrections.
However, it is very difficult to calculate the matrix elements for 
heavy-light systems with HQET. This is because in the heavy-light
system the self-energy correction gives a significant contribution to
the energy so that the noise to signal ratio of the heavy-light meson
correlators grows exponentially as a function of time.
In fact, recent results of $\hat{g}_\infty$ are  
\begin{align}
 & \hat{g}_\infty = 0.51 \pm 0.03_{\text{stat}} \pm 0.11_{\text{sys}}
 &  \text{for} \quad  n_f = 0~\cite{0310050}, \\
 & \hat{g}_\infty = 0.51 \pm 0.10_{\text{stat}}  
 &  \text{for} \quad  n_f = 2~\cite{0510017},
\end{align}
these have about $5\%$ and $15\%$ statistical errors for
quenched and unquenched cases, respectively.
But such accuracies will 
not be sufficient to test new physics.  
Therefore significant improvements for statistical precision in HQET 
are needed.
Fortunately the two techniques to reduce the statistical error are
developed recently, which are the HYP smearing~\cite{0103029} and  
the all-to-all propagators with low mode averaging~\cite{0505023}.
Calculation of $\hat{g}_\infty$ in quenched QCD using these two methods 
was carried out recently 
by Negishi et al.~\cite{0612029}. 
It is found that the statistical accuracy is drastically improved as
\begin{equation}
\hat{g}_\infty = 0.517 (16)_{\rm stat.} \ \ \ \text{for}\ \ \  n_f = 0.
\end{equation}

Our ultimate goal is to extend the above strategies to unquenched
simulation and give precise values of the $B^* B\pi$ coupling 
with $2+1$ flavors in the continuum limit. 
In this report, we present our high precision study of 
the static $B^* B\pi $ coupling in $n_f=2$ unquenched
QCD combining two techniques of 
the HYP smeared link and the all-to-all propagators. 
Our results also suggest that 
these techniques can be useful to precisely calculate other physics
parameters for heavy-light systems.
%%%%%
\section{Lattice observables}

The $B^* B\pi$ coupling can be obtained from the form factor 
at zero recoil which corresponds to the matrix element 
\begin{align}
\langle B^*(p_{B^*},\lambda)|A_i|B(p_B)\rangle 
|_{\overrightarrow{p_{B^*}}=\overrightarrow{p_B}=0}
=(m_B+m_{B^*})A_1(q^2=0)\epsilon^{(\lambda)}_i,
\end{align}
where $A_1(q^2=0)$ is the matrix element of $B$ to $B^*$ 
at zero recoil with axial current 
$A_i \equiv \bar{\psi}\gamma_5 \gamma_i \psi$
and $\lambda$ stands for polarization~\cite{9807032}.
In the static limit,
\begin{align}
\hat{g}_\infty=A_1(q^2=0)
\end{align}
holds, so that $\hat{g}_\infty$ can be evaluated by lattice
calculations from the ratio of the 3-point and 2-point correlation
functions as 
\begin{gather}
\hat{g}_\infty = R(t,t_A) \equiv \frac{C_3(t,t_A)}{C_2(t+t_A)},
\end{gather}
where 
$C_2(t) \equiv \langle 0 | {\cal O}_B(t) {\cal O}^{\dagger}_B(0)|
0 \rangle$, 
$C_3(t,t_A)\equiv \langle 0 | {\cal O}_{B^*_i} (t+t_A) A_i(t_A)
{\cal O}^{\dagger}_B(0)|0 \rangle$ 
 are 2-point and 3-point functions
and ${\cal O}_B$, ${\cal O}_{B^*_i}$ are some interpolation operators 
for the $B$ meson and $B^*$ meson with polarization in the $i$-th direction.
The lattice HQET action in the static limit is defined as 
\begin{equation}
S = \sum_x \bar{h}(x)\frac{1+\gamma_0}{2}
\left[
h(x)-U_4^\dagger(x-\hat{4})h(x-\hat{4}))
\right],
\end{equation}
where $h(x)$ is the heavy quark field. The static quark propagator 
 is obtained by solving the time evolution equation.
As is well known,
HQET propagator is very noisy, and it becomes increasingly 
serious as the continuum limit is approached.
In order to reduce the noise,
the Alpha collaboration studied  HQET action in which the link
variables are smeared  to  suppress the power divergence~\cite{0103029}.
They found that the HYP smearing significantly suppresses the noise of
 the static heavy-light meson. The error is further suppressed by applying
the all-to-all propagator technique developed by the 
 TrinLat collaboration~\cite{0505023}.
We divide the light quark propagator into two parts:  
the low mode part and the high mode part.
The low mode part can be obtained using low
eigenmodes of Hermitian Dirac operator. The high mode part can be
 obtained by the standard random noise methods with time, color, and
 spin dilutions.  
Combining these propagators, we can obtain the 2-point functions for
the heavy-light meson which are averaged all over the spacetime.
Similarly, the 3-point functions can be divided into four parts : 
low-low, low-high, high-low and high-high parts.

\section{Simulation details}

Numerical simulations are carried out on $12^3\times 24$ 
lattices with two flavors of $O(a)$-improved Wilson quarks 
and the Iwasaki gauge action
at $\beta = 1.80$ corresponding to $a^{-1}=0.9177(92){\rm GeV}$. 
We make use of 100 gauge
configurations  provided from
CP-PACS collaboration through JLDG~\cite{0105015}. 
We use the $O(a)$-improved Wilson fermion for light valence quark with
clover coefficient $c_{sw}=1.60$.
In the HQET, we use static action with HYP smeared links with 
smearing parameter values $(\alpha_1,\alpha_2,\alpha_3)=(0.75,0.6,0.3)$.
The $B$ and $B^*$ meson operators are smeared with a function 
$\phi(r)=\exp{(-0.9r)}$.
We obtain the low-lying eigenmodes of the Hermitian Dirac operator
using implicitly restarted Lanczos algorithm.
The low mode correlation functions are
computed with $N_{ev}$ = 200 eigenvectors of the low-lying eigenmodes
of the lattice Hermitian Dirac operator. 
The high mode correlation functions are obtained
by using the quark propagator with the source vector generated by the
complex random $Z_2$ noise.
The number of the random noise and the number of time dilution for
each configuration are set to $N_r=1$ and $N_{t_0}=24$, respectively.
This setup is based on the experience from the work 
by Negishi et al.~\cite{0612029}.

\section{Results}

\begin{figure}[!ht]
\begin{minipage}{.45\linewidth}
%\hspace{-1.25cm}
\rotatebox{0}{
\includegraphics[width=6cm,clip]{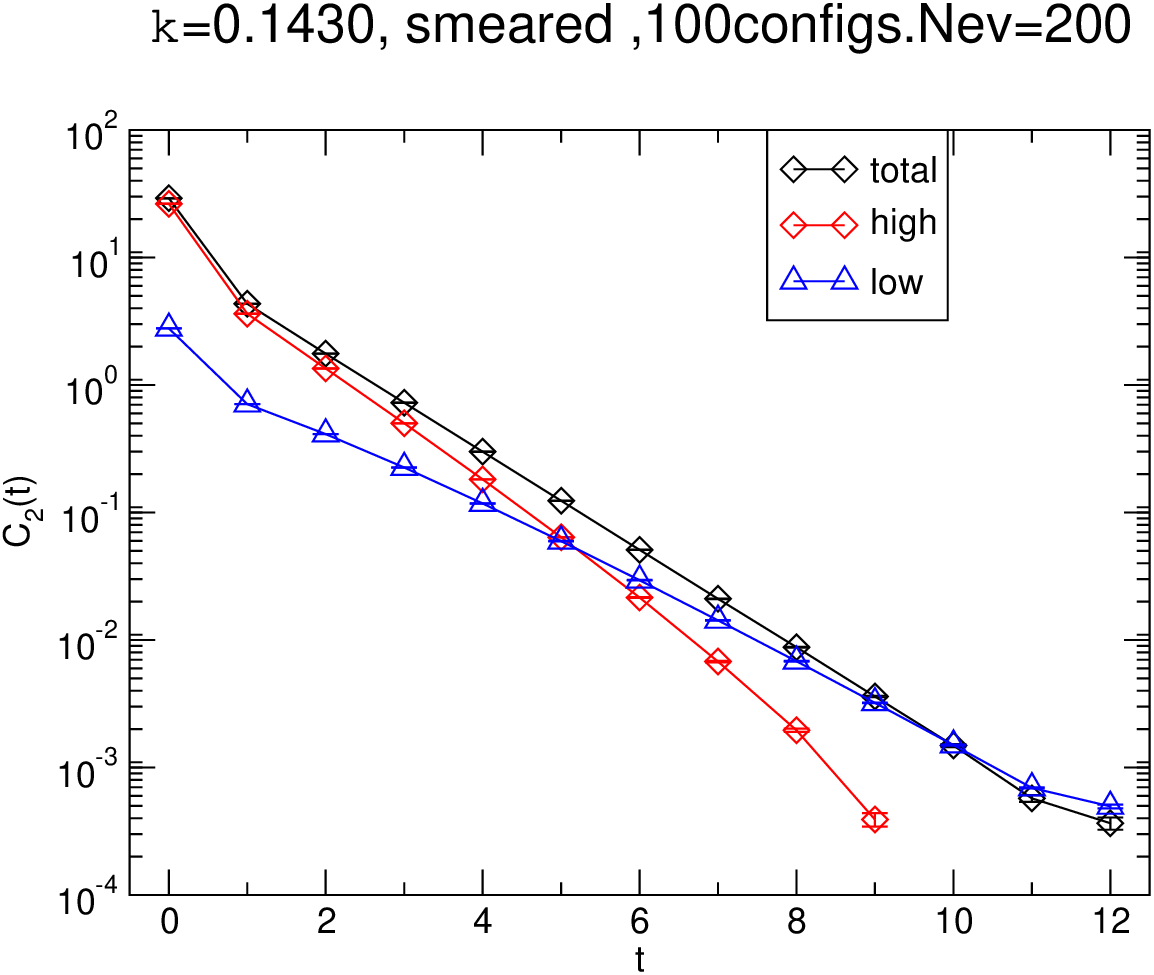}
}
\caption{Low mode and high mode contributions to the 2-point 
functions for $\kappa=0.1430$. 
Blue and red symbols represent 
low and high mode contributions, respectively.
We can see low mode becomes dominant for $t$ larger than five. 
}
\label{fig:2pt}
\end{minipage}
\begin{minipage}{1.65cm}
\
\end{minipage}
\begin{minipage}{.45\linewidth}
\rotatebox{0}{
\includegraphics[width=6cm,clip]{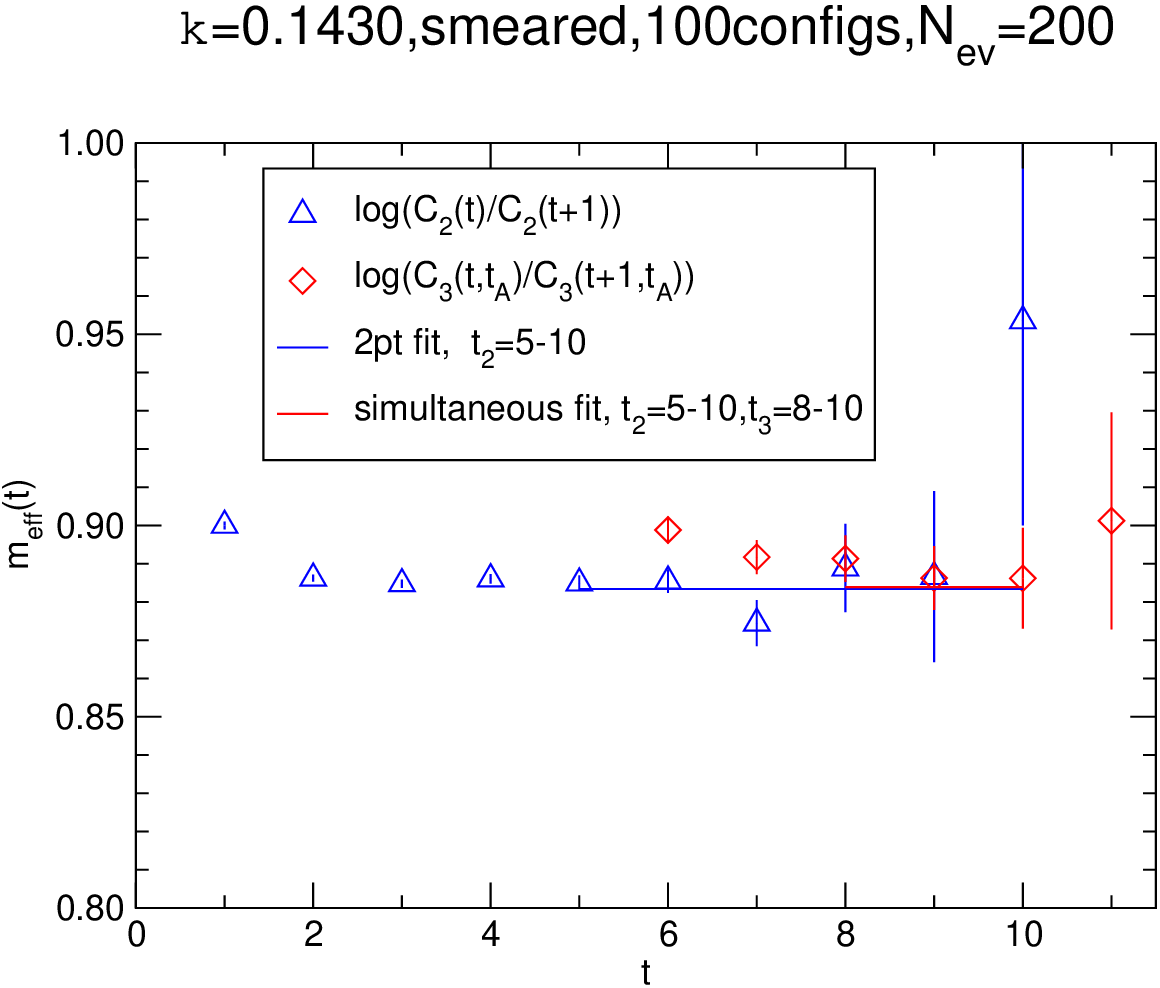}
}
\caption{The effective masses of the 2-point and the 3-point functions
with $\kappa=0.1430$ and $t_A=5$.
Blue line represents the fit for the 2-point functions only.
Red line  corresponds to simultaneous fit 
for the 2-point and 3-point functions.
}
\label{fig:mass}
\end{minipage}
\end{figure}

We show the result of the 2-point correlator with $\kappa=0.1430$ in
Fig.~\ref{fig:2pt}, which displays whether the improved technique
works successfully. Indeed, both the low mode part 
and the  high  mode part have
small statistical errors. The effective mass of the 2-point function
is shown in Fig.~\ref{fig:mass}. We observe a nice plateau 
at  $t \geq  4$. 
From this result we take $t_A$=5 as a reasonable choice of 
the time difference between the current $A_i$ 
and the $B$ meson source.  

%%%%%
%%%%%

\begin{figure}
\begin{minipage}{.45\linewidth}
%\hspace{-1.4cm}
\rotatebox{0}{
\includegraphics[width=6cm,clip]{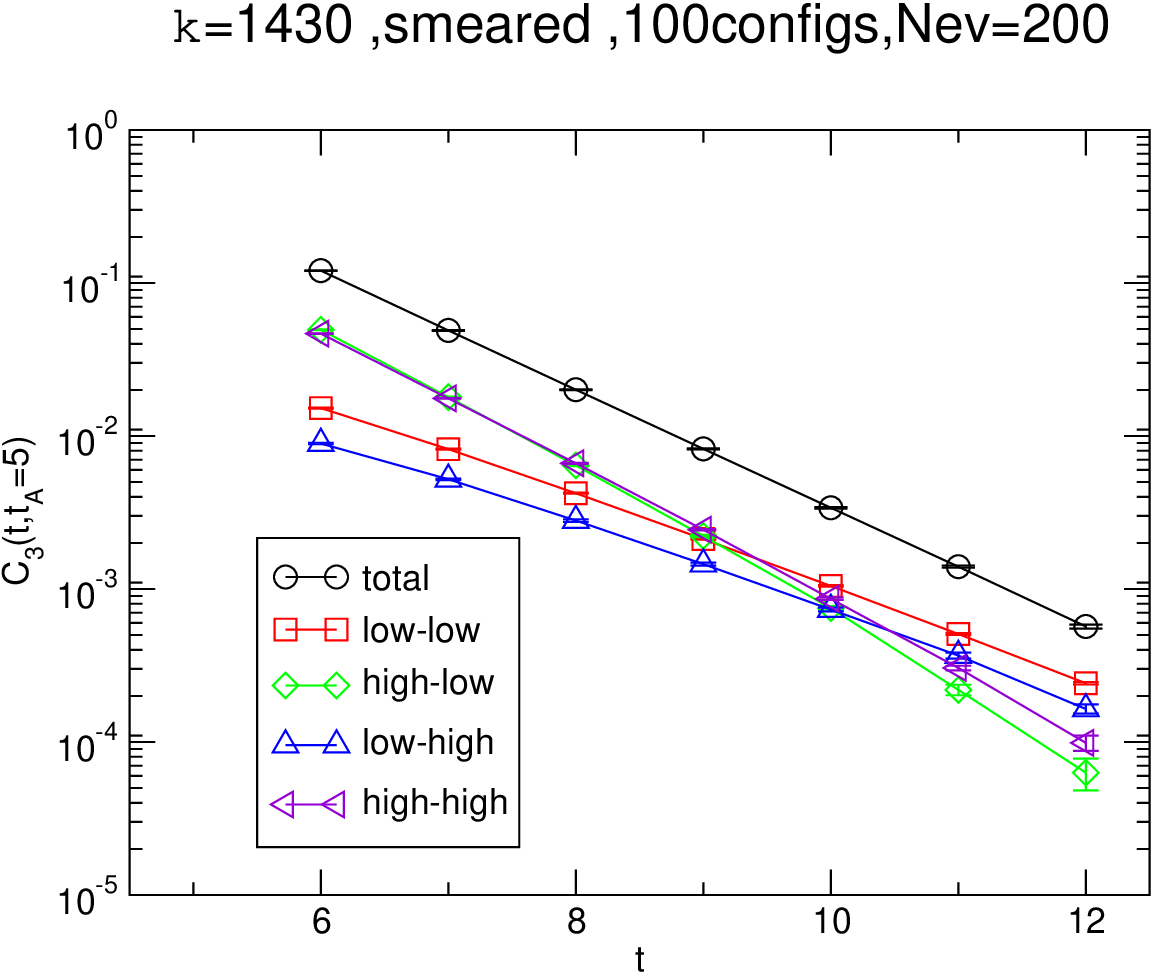}
}
\caption{Contributions of low-low, high-low, low-high and high-high
parts to the 3-point functions  
for $\kappa=0.1430$.}
\label{fig:3pt}
\end{minipage}
\begin{minipage}{1.65cm}
\
\end{minipage}
\begin{minipage}{.45\linewidth}
\rotatebox{0}{
\includegraphics[width=6cm,height=5cm,clip]{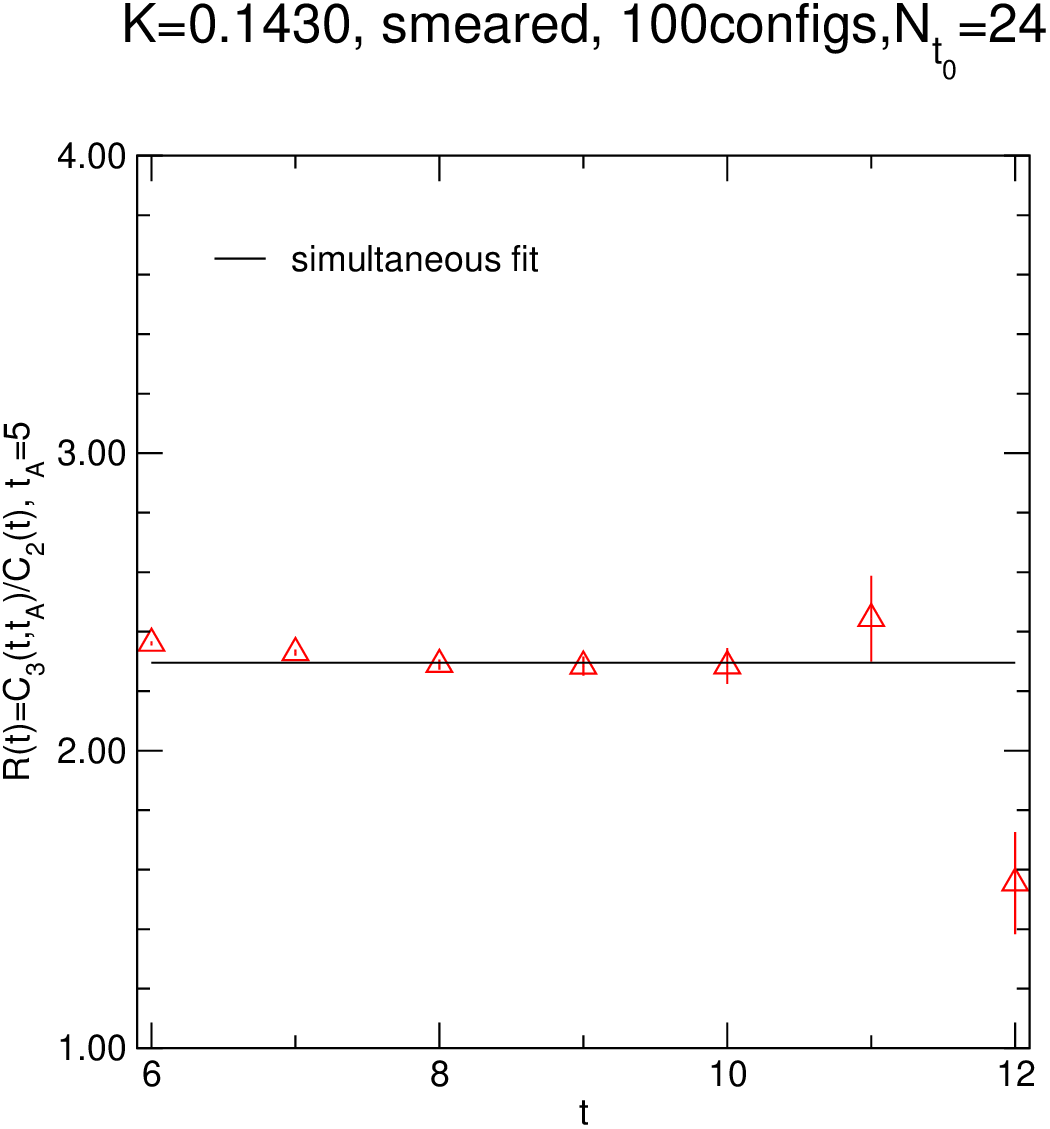}
}
\caption{The ratio of the 3-point and 2-point functions
at each $t$ for $\kappa=0.1430$. The solid line represents the ratio
of  fit parameters $Z_3/Z_2$. 
}
\label{fig:ratio}
\end{minipage}
\end{figure}

Fig.~\ref{fig:3pt} shows the time dependence of the 3-point functions for
$\kappa=0.1430$.
Fluctuations of high-modes for 3-point functions are indeed suppressed,
so we can get a good plateau for the effective mass of 3-point
functions as in Fig~\ref{fig:mass}.  
We use the following fit functions
\begin{equation}
\begin{split} 
C_2(t)=Z_2 \exp{(-mt)}, \\
C_3(t)=Z_3 \exp{(-mt)},
\end{split}
\end{equation}
where $Z_2$ and $Z_3$ are constant parameters and $m$ corresponds to
the heavy-light meson mass. The bare $B^* B\pi$ coupling can be
obtained by the ratio of the fit parameters as 
$\hat{g}_\infty^{\rm bare}=Z_3/Z_2$ .
Fit ranges for the 2-point and the 3-point functions are 
$5\sim 10$ and $8\sim 10$, respectively.
Fig.~\ref{fig:mass}  shows that the effective masses and the fit results are 
all consistent as we expected.

Fig.~\ref{fig:ratio} compares the results of 
the $B^* B\pi$ coupling for
$\kappa=0.1430$ determined from two different methods.
It is found that the ratio $C_3(t)/C_2(t)$ at each $t$ 
and the ratio of the fit parameters $Z_3/Z_2$ 
give consistent value of the $B^* B\pi$ coupling.
We will use the latter result  $Z_3/Z_2$
to determine $\hat{g}_\infty$ in the following analyses.
The physical value of the $B^* B\pi$ coupling is obtained 
by multiplying the renormalization constant. We use the one-loop
result of renormalization factor for the axial vector 
current
\begin{gather}
A_i  = 2\kappa u_0 Z_A 
\left(
1+b_A\frac{m}{u_0} 
\right)
A_i^{lat}, \\
u_0 = 
\left(
1-\frac{0.8412}{\beta}
\right)^{\frac{1}{4}}, \ \ \ \ 
 b_A = 1+ 0.0378 g_{\bar{MS}}^2(\mu), \nonumber
\end{gather}
where the gauge coupling $g_{\bar{MS}}^2(\mu)=3.155$ and 
$Z_A = 0.932$ for $\beta=1.80$ as given in Ref.~\cite{0105015}.
We arrive at our preliminary results of  $\hat{g}_\infty$ 
for our $\kappa$ values in Table 1.
\begin{table}[h]
\label{table:ghat}
\begin{center}
\begin{tabular}{c|l|l|l|l}
$\kappa$ & 0.1409 & 0.1430 & 0.1445 & 0.1464 \\ \hline
$\hat{g}_\infty$         &  $0.612(5)_{\text{stat}}$   & 
$0.598(5)_{\text{stat}}$ &  $0.591(4)_{\text{stat}}$   & 
$0.578(5)_{\text{stat}}$ 
\end{tabular}
\caption{Preliminary results of $\hat{g}_\infty$.} 
\end{center}
\end{table}%
%%%%%

We take the chiral extrapolation of the $B^* B\pi$ 
coupling using the data with  four quark masses.
Employing the following fit functions (a), (b), (c), 
\begin{eqnarray}
{\rm (a)}\ \  g(m_\pi^2)_1  &=& g(0)+A_1 m_\pi^2, \nonumber \\
{\rm (b)}\ \  g(m_\pi^2)_2  &=& g(0)+A_1 m_\pi^2 +A_2 (m_\pi^2)^2, \nonumber \\
(c)\ \  g(m_\pi^2)_3  &=& g(0)\left(
                         1-g(0)^2\frac{7}{64\pi^2}\frac{m_\pi^2}{f_\pi^2}
                         \log{(m_\pi^2)}
                   \right)
                   +A_1 m_\pi^2 +A_2 (m_\pi^2)^2, \nonumber 
\end{eqnarray}
we carry out the linear extrapolation,
the quadratic extrapolation, 
and the quadratic plus chiral log extrapolation where the log
coefficient is determined from ChPT\cite{9312304}.
We use three, four and  four data point for the fits, respectively. 
We obtain physical values of the $B^* B \pi$ coupling in the chiral limit as  
$\hat{g}_\infty = 0.57(1), 0.57(2), 0.52(1) $ 
from the linear fit, the quadratic fit and 
the quadratic plus chiral log fit, respectively.
We take the average of the results from the linear fit and the quadratic plus
chiral log fit as our best value and take half the difference as
the systematic error from the chiral extrapolation.
Other systematic errors are the perturbative error of $O(\alpha^2)$, 
and the discretization error of $O((a\Lambda)^2)$. 
Including these errors estimated by order counting, 
our preliminary result of $\hat{g}_\infty$ is 
\begin{equation}
\hat{g}_\infty^{n_f=2} =  0.55(1)_{\text{stat.}}(3)_{\text{chiral.}}
(3)_{\text{pert.}}(6)_{\text{disc.}} \ \ \text{at}\ \ \beta=1.80.
\end{equation}
We find that the discretization error is dominant 
in our simulation  on this coarse lattice.  %%
%%%%%
\begin{figure}[!ht]
\hspace{3mm}
\vspace{5mm}
\begin{center}
\rotatebox{0}{
\includegraphics[width=7cm,clip]{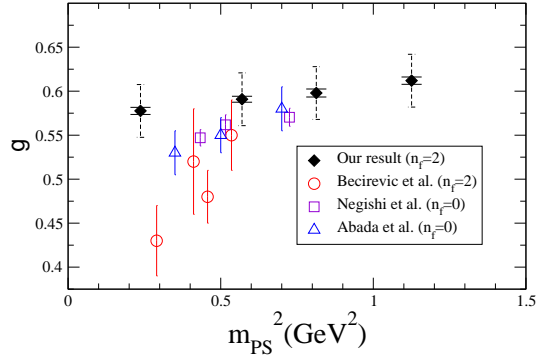}
}
\caption{Comparison with other
calculations~\cite{0310050,0510017,0612029}. 
In our results, small and large errors represents statistics error 
and perturbative error respectively.
}
\label{fig:comparison}
\end{center}
\end{figure}%
%%%%%%

\section{Conclusion}

In this report, we computed the 
$B^* B\pi$ coupling on unquenched lattices 
using the HYP smearing and the all-to-all propagators.
Using the low mode averaging with 200 eigenmodes, the statistical error
becomes tiny for all the quark masses, giving $\sim 2 \% $ in the
chiral limit. However, since the dominant error is from the
discretization  for our simulation on this coarse lattice, 
we need to simulate on finer lattices. 
In Fig.~\ref{fig:comparison},  
we compare the recent results  of the $B^*
B\pi$ coupling~\cite{0310050,0510017,0612029}.
The improvement in statistical precision is drastic, 
which proves the power of the improvement techniques examined in this report. 
This result implies that one can also precisely calculate other quantities 
such as the B meson decay constant or $\hat{g}$ 
with $1/M$ corrections. 
\section*{Acknowledgments}
We would like to thank JLDG for providing with unquenched
configurations from CP-PACS collaboration.  
The numerical calculations were carried out on the vector
supercomputer NEC SX-8 at Yukawa Institute for Theoretical Physics, 
Kyoto University. 
The simulation also owes to a gigabit network SINET3 
supported by National Institute of Informatics,  
for efficient data transfer suppoted by JLDG.
This work is supported in part by the Grant-in-Aid of the
Ministry of Education (Nos. 19540286, 19740160).


\begin{thebibliography}{99}

\bibitem{0612020}
B. Aubert et al.(BABAR Collaboration),
\emph{Measurement of the B0 ---> pi- l+ nu form-factor shape and
branching fraction, and determination of |V(ub)| with a loose neutrino
reconstruction technique},
\emph{Phys.Rev.Lett.} {\bf{98:091801}} (2007), [{\tt hep-ex/0612020}]
 
\bibitem{0310050}
A. Abada, D. Becirevic, Ph. Boucaud, G. Herdoiza, 
J.P. Leroy, A. Le Yaouanc and  O. Pene,
\emph{Lattice measurement of the couplings affine g infinity and g(B*B pi)},
\emph{JHEP}. {\bf{0402:016}} (2004), [{\tt hep-lat/0310050}] 

\bibitem{0510017}
D. Becirevic, B. Blossier, Ph. Boucaud, J. P. Leroy, 
A. LeYaouanc and O. Pene,
\emph{Pionic couplings \^g and \~g in the static heavy quark limit}.
\emph{PoS LAT2005},
 {\bf{212}} (2006), [{\tt hep-lat/0510017}] 

\bibitem{0103029}
A. Hasenfratz and  F. Knechtli,
\emph{Flavor symmetry and the static potential with hypercubic
blocking}
\emph{Phys.Rev.D64}  {\bf{034504}} (2001), [{\tt hep-lat/0103029}] 

\bibitem{0505023}
J. Foley, K. J. Juge, A. O'Cais, M. Peardon,
S. M. Ryan and J. I. Skullerud,
\emph{Practical all-to-all propagators for lattice QCD},
\emph{Comput.Phys.Commun}. {\bf{172:145-162}} (2005),
 [{\tt hep-lat/0505023}] 

\bibitem{0612029}
S. Negishi, H.Matsufuru, T. Onogi,
\emph{Precision study of B* B pi coupling for the static heavy-light
meson},
\emph{Prog.Theor.Phys.} {\bf{117:275-303}} (2007), [{\tt hep-lat/0612029}] 

\bibitem{9807032}
G. M. de Dibitiis, L. Del Debbio, M. Di Pierro, J. M. Flynn,
C. Michael and J. Peisa(UKQCD Collaboration),
\emph{Towards a lattice determination of the B* B pi coupling},
\emph{JHEP}. {\bf{9810:010}} (1998), [{\tt  hep-lat/9807032}] 

\bibitem{0105015}
A. Ali Khan et al. (CP-PACS Collaboration)
\emph{Light hadron spectroscopy with two flavors of dynamical quarks
on the lattice},
\emph{Phys.Rev}. {\bf{D65:054505}} (2002), 
\emph{Erratum-ibid}. {\bf{D67:059901}} (2003),
[{\tt hep-lat/0105015}] 

\bibitem{9312304}
H.Y. Cheng, C.Y. Cheung, G.L. Lin, Y.C. Lin, T.M. Yan and H.L. Yu, 
\emph{Corrections to chiral dynamics of heavy hadrons: 
SU(3) symmetrybreaking},
\emph{Phys.Rev.} D{\bf{49}}5857-5881 (1994),
\emph{Erratum-ibid.} D{\bf{55}}:5851-5852 (1997), 
[{\tt  hep-ph/9312304}] 

\end{thebibliography}
\end{document}